# Attacks on Local Searching Tools

Seth Nielson  
*seth@sethnielson.com*

Seth J. Fogarty  
*sfogarty@cs.rice.edu*

Dan S. Wallach  
*dwallach@cs.rice.edu*

*Department of Computer Science, Rice University*

## 1 Introduction

The *Google Desktop Search* is an indexing tool, currently in beta testing, designed to allow users fast, intuitive, searching for local files. The principle interface is provided through a local web server which supports an interface similar to Google.com's normal web page. Indexing of local files occurs when the system is idle, and understands a number of common file types. A optional feature is that Google Desktop can integrate a short summary of a local search results with Google.com web searches. This summary includes 30-40 character snippets of local files.

Despite the obvious usefulness of local searching, there has been considerable discussion about the privacy implications of local indexing systems like the Google Desktop Search. Fundamentally, indexing systems make data access easier for legitimate users without also making it easier for unauthorized individuals. Additionally, some do not consider the authors of such software to be trusted parties and fear that the applications might leak information to corporate entities.

Google Desktop Search was created and is marketed with these security concerns in mind. Not only are a number of security features evident in this tool, but the online literature provided by Google is emphatic about the safety of a user's private information. It should be noted that the Google Desktop Search is designed to be used on single-user machines running Windows; as an administrative process it can, and does, index all files regardless of owner. Obviously this is unacceptable on multi-user machines.

In our research we searched for a vulnerability that would release private local data to an unauthorized remote entity. Our focus was on the small snippets of local data that the integration feature handled. We realized that this feature was combining local private data with remote public data in a possibly unsafe environment. We present two different attacks that exploit this vulnerability.

The remainder of this report is structured as follows. In the second section, we describe the operation of the Google Desktop Search in greater detail. In the third and fourth sections, we describe our attacks on the integration feature. The fifth section presents our analysis of the situation and discusses possible solutions as well as the solution chosen by Google.

Figure 1: Google Desktop's integration of local search results into web searches.

## 2  Background on Google Desktop Search

After installation on a local host, the Google Desktop application begins indexing local files. The current beta version indexes a variety of common files, including Microsoft Word, Excel, and PowerPoint, email stored in Outlook or Outlook Express and AOL instant messaging conversations [3]. The indexing process generally consumes system resources only when the system is idle.

Queries against the local search index are performed through a web interface. The Google Desktop application installs a local web server and the user interface is provided by web pages served by this internal server. The layouts of the desktop search and the returned results pages are almost identical in style and form to the Google.com web search and results pages.

One unusual feature of the Google Desktop application is the integration of local result snippets into remote Google.com web searches. If this option is enabled, when a user performs a search at Google.com, his or her results page is modified to display a small number of matching results on the local computer along with very short, 30-50 character, snippets of the matching text. Figure 1 shows an example of this feature. When the user performed a search for "foo," a single web page presented the results of local searches, product searches, sponsored advertisements, and normal web searches. This level of search integration allows the user to find what he or she is looking for without making each of those queries individually.

### 2.1  Privacy objectives

Google states that "We treat your privacy with the utmost respect. The Google Desktop Search program does not make your computer's content accessible to Google or anyone else" [3]. The privacy policy elaborates, "Your computer's content is not made accessible to Google or anyone else without your explicit permission" [4].

To protect the Google Desktop application from being used for malicious purposes, it was designed to be inaccessible to remote users. The local web server only accepts connections *to* localhost or `127.0.0.1`, regardless of the source address. By ignoring the source address and accepting packets based strictly on the destination address the web-server sidesteps the problem of source-address forging. Only network connections originating on and connecting to the local machine will ever be seen as connecting to localhost. This design clearly and elegantly prevents external computers from directly querying the local web server.

## 2.2 Desktop integration

In our study, we decided to further examine the local search integration feature of the Google Desktop. We wanted to learn how this feature operated, as it seemed a promising avenue of attack. If an external attacker could read the local search integration results, significant private information would be leaked. If the attacker could choose the search terms, this attack could be particularly damaging to users, particularly those who keep sensitive information, such as passwords or credit card numbers, stored in their private files.

According to Google,

> Desktop Search allows you to simultaneously send your query to two different programs and locations. One query goes to Google, which performs a standard Google Web Search. A duplicate query goes to the Desktop Search application running on your computer, which searches the information the application has indexed for you. Desktop Search intercepts Google's results page before you see it and adds your Desktop Search results just above your web results so you can see both at once. [5]

The raw HTML seems consistent with this explanation. The local search results appear to be a normal part of the file. No JavaScript, frames, or other directives to the web browser were used to integrate the local results with the web results. To gain a better understanding of how the integration worked, we conducted a number of experiments.

### 2.2.1 Network sniffing

Our first experiment with the integration feature was to simply capture the network packets of a Google query. This could determine if any local information went over the network. We used Ethereal[1] to monitor communication between our local computer and Google. We captured traces of Google searches on two different computers, one on Rice's network and one on a private DSL connection at home. Likewise, we captured plain text sessions (HTTP/1.0) and gziped (HTTP/1.1) sessions. We discovered that:

1. The Google Desktop Search does not transmit private data during a web search.
2. None of the response packets from Google captured by Ethereal had any of the integration data.

---
[1] Available at `http://www.ethereal.com`

In other words, we verified that integration is indeed a local operation. We now knew that some agent was running locally on our machine that would intercept incoming Google result pages and integrate the results from local indexing. Two questions remained. We were not certain where exactly integration happened, and we did not know what triggered the local search.

### 2.2.2 Replay reconnaissance

Google's explanation of local result integration seemed to suggest that the integration was initiated by the originating page, sending off two different requests. Ethereal cannot detect packets sent on the loopback interface, so that remained a possibility. Two other logical options were that it was triggered by the outgoing request or the incoming response.

To determine the answer, we used Ethereal to save several Google packet traces to disk. These packets contained the response to a previous Google search for some term $Search_A$. To feed these back to the browser, we wrote a simple Python script that would accept HTTP proxy requests but would then always replay the packets that we had previously captured.

We then opened a browser and configured it to use our proxy server. We pointed the browser to `www.google.com`, and entered a different search ($Search_B$). As a result, the proxy server returns the results for $Search_A$, but the local search results for $Search_B$ were integrated with the web page. Thus, the local search engine only considers the outgoing request, not the inbound response, for determining its own search query. We also observed that the integrated results seemed to be inserted directly after the second TCP response packet, thus simplifying the matter of modifying packets and dealing with the compressed HTTP/1.1 data stream.

### 2.2.3 Socket scrutiny

We knew where local search results are integrated, and we knew where the local search engine got its query. However, we needed to know what conditions were necessary to trigger this process. We were curious if any HTTP request to Google by any program running on the user's machine would trigger integration, or if the request need originate from the main Google.com web page. Likewise, we were curious whether the integration would occur even when a web browser was configured to use a proxy server rather than directly connecting to Google over the Internet. To test this, we wrote a simple Python script that would open a socket to Google.com and execute a search request. We similarly attempted submitting a request through a proxy. We found that both the direct request and the proxied request would both have local search results integrated. From this, we conclude

1. Google Desktop must be observing all outgoing network connections.

2. Google Desktop performs packet analysis to identify HTTP proxy connections in addition to looking for direct connections to Google.

3. The search requests did not need to originate from a web browser visiting Google.com.

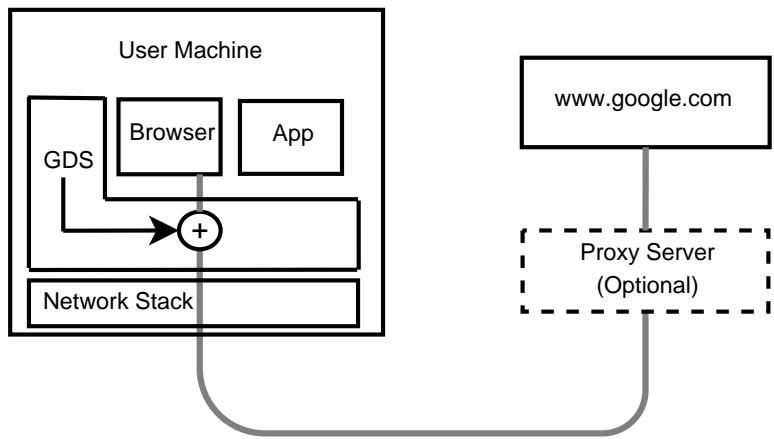

Figure 2: Normal operation of Google Desktop Search. GDS intercepts all outbound network connections and integrates local search results with Google web queries.

4. Integration is triggered by observing outgoing packets, and occurs after packets are received, but before they are given to the web browser or application.

Figure 2 describes our understanding of this process.

The challenge, from the perspective of an attacker, is to make integration-triggering network connections *from* the target's computer and to read the results after integration has occurred.

## 3  Java applet attacks

Because the Google Desktop application bases its decision to integrate strictly on network traffic, all that is required for an eavesdropper to obtain an integrated web page is to open a socket on the target computer and send an HTTP request to Google.com, either directly or through any server configured as a web proxy server. This is well within the capabilities of a Java applet, even when running with the restrictive "sandbox" security model.

Downloading and running a Java applet is an automated process for most Java-enabled web browsers. Thus, any web page a victim loads off of a hostile server may include a malicious Java applet. This applet will be downloaded and executed inside the web browser without user intervention. Such Java applets are normally used to implement a variety of features, not available through regular HTML and JavaScript, that range from user interface widgets to complex games and animations.

The Java "sandbox" security model places a number of restrictions on untrusted Java applets that ensure they are safe to run. In particular, applets are not allowed to read or write any local files, nor are they allowed to make network connections to any host beyond the one they originated from. While numerous security holes have been found in Java [2], we do not exploit any of these holes to effect an attack. Once the malicious applet starts running, the attacker can use it to make queries against the local search engine until the entire browser application is closed. These queries will return the snippets of text the integration feature provides.

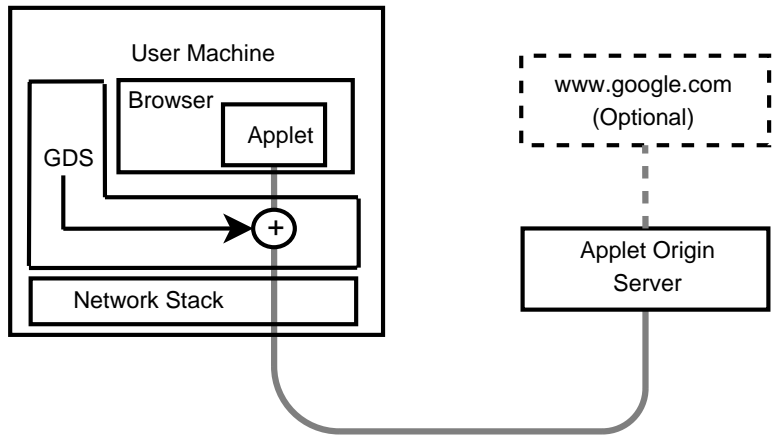

Figure 3: A Java applet, legally connecting to its origin server, can fool the Google Desktop service into integrating local search results into non-Google pages.

### 3.1 Implementation

To accomplish the attack, we took advantage of the Java applet's ability to connect to the machine from which it is loaded. That machine, under the control of the attacker, can run a web proxy server of the attacker's design. The applet can legally connect to the proxy server and make requests for the proxy to fetch results from Google.com. The proxy can return the results from any previous Google page, as those results will not actually be used. The Google Desktop's local search integration cannot distinguish this connection from the Java applet with a legitimate connection from a web browser, and will thus integrate the search results where they can be read by the applet. Of course, the applet can subsequently transmit these results back to the server from which it was loaded. We diagram this process in Figure 3.

In our implementation, we designed the applet to first open a control channel with the server. This allows the server to issue search queries to the applet. When applet receives one of these queries, it connects back to the server, as described above, to make a proxy request and subsequently capture the local search results. These results are returned to the server over the control channel. This gives the attacker real-time control over the applet, allowing him to try a number of different queries and to refine them with the results of the earlier queries. Thus the attacker can search interactively for sensitive, private, information on the target computer.

### 3.2 Feasibility

The main impediment to performing this attack is to somehow trick the user of the target computer to visit a hostile web site. Of course, the attacker could break into (i.e., deface) a legitimate web site that the target user regularly visits. Likewise, the attacker could perform some kind of social engineering, perhaps with spam-like email advertisements, to entice the target user to visit the hostile site.

Once the target user loads the hostile page, the damage has been done. There is no need for the attacker to be on the same network as the target, nor is there any need for the attacker to "break into" the target machine in any traditional fashion. Furthermore, because all of the interaction between the target machine and the attacker uses standard web traffic, most commercial firewalls would offer no protection against this attack. Of course, if the user has disabled Java, or the firewall has filtered out any Java applets, then this attack would fail, although similar attacks might be possible with other programmable content types like Macromedia's Flash. On the other hand, if the user has disabled local search integration with web searches, the attack would be completely defeated. This only requires deselecting a single checkbox on the "preferences" screen.

### 3.3 Man-in-the-middle variants

In many cases, an attacker will be in a position to observe the network traffic coming from the target's computer and can inject network traffic that pretends to come from Google or any other network host. Such "man-in-the-middle" attacks are particularly easy to perform when the attacker and target are sharing the same 802.11 wireless network. These networks are increasingly available in many hotels, airports, and cafés and do not use any 802.11 security features such as WEP encryption. Even on a private network with WEP encryption, an attacker could easily break the encryption [11].

The attacker's goal, in such a scenario, will be to trick the target computer's web browser into loading the attack applet within an unrelated web page. This takes advantage of a common practice, particularly with web advertising, where web sites will include Java applets or Flash animations hosted by third parties. While a number of techniques may be used to accomplish the attack, probably the simplest is to passively read every web page loaded by the target, looking for references to external applets. Upon seeing this, the attacker can predict that the target will make a DNS lookup for the applet host. The attacker then issues a DNS response that maps that DNS name to the attacker's IP address. Eventually, the correct DNS response will arrive, but the target machine will discard it and fetch the applet from the attacker. A sophisticated attack applet could be engineered to impersonate the original applet; the attack could even be implemented as a virus attached to the original applet.

While web sites could take countermeasures to defeat this attack, such as operating entirely with SSL/TLS encryption and authentication, this seems unlikely to be widely adopted. Instead, users of wireless networks could tunnel all of their traffic through a virtual private network (VPN). VPN systems are generally provided by corporations to allow traveling users to access the company's intranet while traveling outside. VPN technologies would defeat man-in-the-middle opportunities on the target's local wireless connection. Unfortunately, VPNs are generally only available to corporate users. Furthermore, some wireless systems restrict the ports where they will carry traffic, sometimes interfering with some VPN systems.

### 4 JavaScript-based attacks

In many organizations, the use of Java or other generally programmable plugin systems like Macromedia's Flash is forbidden. Such organizations are uncomfortable with the risk that a crafty attacker could circumvent the protections enforced by these tools. By banning these tools, a possible vector of attacks is removed. In practice, while this might degrade the experience of many web sites, most will continue to operate cor-rectly. Because many users do not have Java or Flash installed at all, even the most multimedia-laden web sites will commonly offer a simplified, plain HTML view. In contrast, the JavaScript scripting language is used extensively by numerous web sites and is widely supported by commercial browsers. Disabling JavaScript renders many such web sites unusable. As such, it would be valuable, from the perspective of an attacker, to discover an attack that need not rely on Java.

### 4.1 Ley's attack

Ley recently described an attack that takes advantage of Google's web customization features [7]. For web sites that use Google to "power" their site searches, Google provides an interface for sites to add their logos and such to the Google results. Ley used this feature to inject malicious JavaScript into a Google web page which would implement a "phishing" attack. Similarly, this inserted JavaScript could also be used to read other contents of a Google web page, including any integrated local search results, and send them to a third party. Google's web servers now filter out any JavaScript or VBScript references passed through the customization interface, thus defeating Ley's attack.

### 4.2 Man-in-the-middle variants

Despite Google having closed Ley's security hole, if an attacker is appropriately positioned in the network to perform a man-in-the-middle attack (see Section 3.3), the attacker should be able to modify any page transmitted from Google to include malicious JavaScript. The Document Object Model (DOM) allows scripts of this kind to fully traverse and extract all elements of the HTML page. A script can return the extracted elements to the attacker in a number of ways, including passing them as arguments to a CGI script on a colluding web server.

In the same fashion that our Java attack used forged DNS results to strategically redirect queries from legitimate web servers, we could similar interpose when a target host's web browser does a DNS lookup on `www.google.com`, redirecting the target to the attacker's machine. The attacker would then dispatch the query to the real Google server, add in some malicious JavaScript, and pass the results to the target.

The malicious JavaScript, as in the Ley attack, would read the contents of the local search results, and transmit the results back to the attacker, perhaps by opening a zero-height internal frame (IFRAME) that would not be visible to the user. Furthermore, the page returned by the attacker could easily be redirected to perform another Google query. The attacker would thus be able to make interactive queries against the local search service without the target machine's user being aware of the attack.

### 4.3 Implementation

We designed a "proof-of-concept" of this attack to study its viability. Using our wireless network, we programmed a laptop to listen for DNS requests for `www.google.com` and respond with the IP address of the laptop. This places the attacker in the position of being a (transparent) proxy server between the target and Google. We had the proxy insert JavaScript which would attempt to read the integrated local search results.

Our JavaScript successfully read the local search results and reprinted it at the bottom of the web page as proof that we could, indeed, extract the local results. We felt it was unnecessary, for the proof-of-concept, to transmit it back to the attacker. (The open-source DNS hijacking tool we used, *dnshijacker*[2], has been ported to numerous platforms and is very easy to use.)

### 4.4 Feasibility

Google appears to have fixed Ley's JavaScript vulnerability, but the man-in-the-middle attack is still entirely feasible, particularly when the target computer is using a wireless network and the attacker is physically nearby.

Making the attack interactive is much more complicated and more likely to be detected. To be interactive, the JavaScript must pop-up or "pop-under" another window that allows the attacker some modicum of control over Google searches (although this might be successfully hidden in a zero-height IFRAME). For the attacker's server to instruct the target machine what search to perform next, the JavaScript must poll for new instructions, perhaps by refreshing at regular intervals. Such behavior might increase the likelihood that the attack is detected.

As described in Section 3.3, users and web sites can take countermeasures to reduce their exposure to man-in-the-middle attacks. If Google, for example, were to offer all of its services over SSL/TLS and users exclusively visited `https://www.google.com` rather than the `http` version, the man-in-the-middle would be unable to put the hostile JavaScript into the web page.

## 5  Attack analysis

Both versions of our attack, whether using Java or JavaScript, take advantage of *composition* effects. Java, by itself, has a security policy that gives applets a limited ability to make network connections. When used in traditional web pages, this allows applets to have useful behaviors without compromising a user's security. Likewise, the Google Desktop's local search integration feature, by itself, injects local search results into web queries, giving users an improved search experience. So long as network connections are only coming from "trusted" sources, like a web browser, there is no danger of the local search results being leaked. However, when an attacker composes these two systems, the attacker can use a property of applet security to help violate a property of the Google Desktop's security. Such composition effects are one of the most difficult issues in the engineering of secure software; an attacker need only find a single unusual combination of features to accomplish an attack, while the system engineer must consider all possible combinations to prevent any possible attack. As a result, the attacker has a significant advantage. The traditional response is to engineer systems in a conservative fashion, using simple, mature mechanisms. The Google Desktop's use of low-level mechanisms to intercept network connections is an example of an "unusual" approach that may be expected to have unintended consequences. A more conservative approach would be to simply keep local search results entirely separate from web search results.

---

[2] http://pedram.redhive.org/projects.php

## 5.1 Proposed solutions

We present five basic solutions that would prevent these and similar attacks. Some of these solutions are necessarily more thorough than others, and not all are feasible.

1. **Not including snippets:** The Google Desktop search system currently integrates snippets of the contents of documents that match the search query. By removing these snippets, perhaps only listing the file names, significantly less information would be available for an attacker. Of course, significantly less value would be present for the user.

2. **Not integrating:** Local search integration is not a fundamental part of the Google Desktop. It is an optional feature that can be disabled by selecting one checkbox in the preferences dialog.

3. **Images:** Instead of inserting text directly into the web page, the Google Desktop could instead insert a reference to an external image, hosted by the Google Desktop's internal web server. A Java applet would only be able to read the name of the image. Malicious JavaScript would likewise only see the name of the image. An attacker would be unable to see the pixels of the image. Unfortunately, such images would not get larger if the user requested larger fonts, nor would they be legible to users with screen-reading software nor would they support cutting and pasting the text within them.

4. **FRAMEs or IFRAMEs:** Rather than inserting the local search results directly into the Google search result, the Google Desktop could insert some HTML that creates an internal frame (IFRAME) element which loads its content from the Google Desktop's internal web server. This IFRAME would have a different "source" than the web page that surrounds it, meaning that hostile JavaScript, even in the main Google page, would be unable to read the local search results.

5. **SOCKS or other proxy styles:** The Google Desktop currently intercepts TCP connects at a low level in the operating system. This could be replaced, perhaps, by explicitly setting a proxy or SOCKS server in the web browser's Internet settings. Of course, if such proxies are already in use, integrating the Google Desktop would be more complicated. Furthermore, while the Java applet attack might be defeated with such settings, the JavaScript attacks would continue to work.

Google decided to follow the IFRAME approach in their new implementation. For contrast, Microsoft's recent MSN Desktop Search[3] appears to have no search integration whatsoever, although it does have an ActiveX control that might be worthy of further investigation.

## 5.2 Security of IFRAMEs

The IFRAME solution completely resolves both the Java applet and JavaScript-based attacks described in this paper. A Java applet, making a simulated Google query, would only see the HTML code to built the IFRAME (`<iframe src="http://127.0.0.1:4664/search?q=foo...">`) but would not be able to see the contents of the IFRAME. Likewise, JavaScript is restricted by the "same-origin" policy, which generally denies scripts from one website the power to access, modify, or manipulate properties of another website originating from a different server [9]. These restrictions also apply to the contents of FRAME and IFRAME elements of HTML pages. The main page's origin is Google.com while the IFRAME's origin is `127.0.0.1`. As a result, even if an attacker injects malicious JavaScript into a Google page, it will not be able to learn the results of local searches.

---
[3]`http://beta.search.msn.com`

One possible method for breaking the same-origin policy would be to use cross-site scripting attacks. These attacks insert JavaScript into target pages by passing the JavaScripts as arguments to web CGI scripts. The Ley attack (described in Section 4.1) is an example of a cross-site scripting attack. Similar attacks have been done against web-mail systems [6] and web single-signon systems [10]. The normal solution is for the intermediary system to aggressively filter out JavaScript and any other active content [1]. Following the Ley attack, Google now does this filtering at Google.com. There does not appear to be any opportunity to mount an attack of this kind against the Google Desktop's local web server. As a result, we believe that the Google Desktop is not vulnerable to this class of attack.

Today, the security of the Google Desktop system is resting on JavaScript's "same-origin" policy. If an attacker can somehow violate this policy, far more serious attacks than merely reading local search results will become possible [8]. As long as users are running reasonably modern web browsers, they should be safe against this class of attacks.

# 6  Conclusions

We found that the Google Desktop personal search engine contained serious security flaws that would allow a third party to read the search result summaries that are embedded in normal Google web searches by the local search engine. While an attacker would not be able to read the victim's files directly, the search results often contain snippets of the file results that will be visible to the attacker. If the victim had a file with a list of web passwords, for example, an attacker might be able to read some of those passwords. These attacks, now fixed by Google, represent a common example of a composition attack, where the attacker could combine unrelated features of the system to violate the security assumptions of a critical service.